\documentclass[twocolumn]{aastex631}

\usepackage{amsmath}
\usepackage{hyperref}

\def\ve#1{\mbox{\boldmath$#1$}}
\def\muas#1{$#1\,\mu\mathrm{as}$}
\def\muasy#1{$#1\,\mu\mathrm{as}\,\mathrm{yr}^{-1}$}

\begin{document}

\title{Improving source positions in the OCARS catalog: A first approach}
\author{Zinovy Malkin}
\affiliation{Pulkovo Observatory, St.~Petersburg 196140, Russia}
\email{zmalkin@zmalkin.com}

\begin{abstract}
OCARS (Optical Characteristics of Astrometric Radio Sources) is a compiled catalog
of various additional data associated with astrometric radio sources whose coordinates
have been determined from very long baseline interferometry (VLBI) observations.
It contains source coordinates, object type, redshift, optical and near-infrared magnitudes.
Until now, OCARS source coordinates were simply copied from input catalogs and,
as a result, were systematically inhomogeneous.
This work is the first attempt to obtain a unified set of radio source coordinates
aligned to the International Celestial Reference Frame (ICRF), more specically to
the third ICRF release, ICRF3.
Comparison of the source coordinates in the old OCARS version as of December 2024
and the new OCARS version as of March 2025 with the ICRF3-SX catalog using
the vector spherical harmonics (VSH) technique showed almost complete elimination
of systematic errors in new OCARS positions relative to the ICRF3 frame.
\end{abstract}


%%%%%%%%%%%%%%%%%%%%%%%%%%%%%%%%%%%%%%%%%%%%%%%%%%%%%%%%%%%%%%%%%%%%%%%%%%%%%%%%%%%%%%%%%%%%%%%%%%%%%%%%%%%%%%

\section{Introduction}

Information about the physical characteristics of radio sources with VLBI-derived
high-precision positions is important for many studies in astronomy, cosmology
and other related fields.
To facilitate this task, the OCARS (Optical Characteristics of Astrometric
Radio Sources) catalog was created almost 20 years ago \citep{Malkin2008}
with the main goal of collecting in one place selected additional data related
to astrometric and geodetic radio sources found in various databases and literature.
This data include source type, redshift, visual and near-infrared (NIR) magnitudes.
Since the release of the first version of OCARS, the catalog has been
continuously evolving by increasing the number of sources, as well as
adding new and refining existing observational data.
The last detailed description of the OCARS catalog was given in \cite{Malkin2018}.
The actual version of OCARS is freely available at the website of the Pulkovo
Observatory\footnote{\url{http://www.gaoran.ru/english/as/ac_vlbi/}}.
A detailed description of the current state of the OCARS catalog is in preparation.

Until recently, OCARS was mostly used as the source of the most complete and
up-to-date data on redshifts and optical magnitudes of astrometric radio sources.
The OCARS source coordinates were just copied from various catalogs
without critical analysis.
In particular, most OCARS source positions were taken from catalogs that
were not properly aligned to the latest implementation in radio of
the International Celestial Reference System (ICRS) endorsed by the IAU, which is currently
the third version of the International Celestial Reference System, ICRF3,
adopted as a new ICRS realization from January 1, 2019 by Resolution
B2\footnote{\url{https://www.iau.org/static/resolutions/IAU2018_ResolB2_English.pdf}}
of the IAU XXXth General Assembly \citep{Charlot2020}.

As a consequence, the OCARS source position system has been significantly inhomogeneous to date.
Given that the number of sources in OCARS significantly exceeds the number of sources
in the ICRF3 catalog, the task of refining the OCARS source coordinates arose, primarily
of a systemic nature, to provide a more homogeneous set of radio source coordinates
in the ICRS/ICRF3 system for various scientific applications.

This paper presents the first results of this work.
The procedure and results of deriving new OCARS source coordinates
is described in Section~\ref{sect:new_positions}.
The details of the vector spherical harmonics technique (VSH decomposition)
used in this study are given in Appendix.

%%%%%%%%%%%%%%%%%%%%%%%%%%%%%%%%%%%%%%%%%%%%%%%%%%%%%%%%%%%%%%%%%%%%%%%%%%%%%%%%%%%%%%%%%%%%%%%%%%%%%%%%%%%%%%

\section{OCARS source coordinates}
\label{sect:new_positions}

The OCARS source list has been composed from several data sources:
\begin{itemize}
\item celestial reference frame (CRF) solutions computed by the
  International VLBI Service for Geodesy and Astrometry \citep{Nothnagel2017} analysis centers
  and mainly using data stored in the IVS data center\footnote{\url{https://cddis.nasa.gov/archive/vlbi/ivsdata/}}:
  AUS catalog (Geoscience Australia, Australia),
  BKG catalog (Federal Agency for Cartography and Geodesy, Germany),
  OPA catalog (Paris Observatory, France),
  USN catalog (U.S. Naval Observatory, USA),
  VIE catalog (Vienna University of Technology, Austria),
  available either through the IVS data center or through institutional depositories;
\item catalogs of radio source positions obtained from observing programs not associated with the IVS;
\item positions of selected radio sources obtained in special experiments.
\end{itemize}

All input catalogs and other data sources contributed to the OCARS
are listed in Table~\ref{tab:ocars_catalogs}, where for each catalog,
the number of sources used in OCARS and the median errors
(coordinate uncertainties reported by the authors)
in R.A. (multiplied by $\cos\delta$, denoted $\alpha^{\ast}$)
and decl., and the median value of the semi-major axis
of the error ellipse are given.
It should be emphasized that the errors for most of
the catalogs listed in Table~\ref{tab:ocars_catalogs} may not
be representative of the accuracy of these catalogs as a whole,
these errors only apply to the subset of sources used in OCARS.

\begin{table*}
\begin{center}
\caption{Source position catalogs that have contributed to OCARS.}
\label{tab:ocars_catalogs}
\begin{tabular}{lrccrccc}
\hline
\hline
Catalog, reference & \multicolumn{3}{c}{Old OCARS (Dec 2024)} & \multicolumn{4}{c}{New OCARS (Mar 2025)} \\
& Nsou & \multicolumn{2}{c}{Median uncertainty, mas} & Nsou & \multicolumn{3}{c}{Median uncertainty, mas} \\
&& $\alpha^{\ast}$ & $\delta$ && $\alpha^{\ast}$ & $\delta$ & s-maj.ax.\\
\hline
USN CRF soluiton (1,2)           & 5789 & ~0.069 & ~0.125 &  5825 & ~0.108 & ~0.191 & ~0.195 \\
RFC, \citet{Petrov2025_RFC}      &  --- &        &        & 16108 & ~0.917 & ~1.810 & ~1.907 \\
WFCS, \citet{Petrov2021_WFCS}    & 5882 & ~0.980 & ~1.920 &   --- &        &        &        \\
LCS2, \citet{Petrov2019_LCS2}    &  849 & ~3.509 & ~3.000 &   --- &        &        &        \\
VIPS, \citet{Petrov2011_VIPS}    &  363 & ~0.441 & ~0.480 &   --- &        &        &        \\
VEPS-1, \citet{Shu2017_VEPS1}    &  150 & ~2.500 & ~4.555 &     8 & ~7.361 & 24.335 & 25.168 \\
OBRS-2, \citet{Petrov2013_OBRS2} &  112 & ~2.441 & ~4.575 &   --- &        &        &        \\
EGaPS, \citet{Petrov2012_EGaPS}  &  103 & ~8.617 & ~5.890 &   --- &        &        &        \\
BeSSeL, \citet{Immer2011_BeSSeL} &   76 & ~5.385 & ~7.150 &     3 & ~3.423 & ~5.900 & ~5.900 \\
VGaPS, \citet{Petrov2011_VGaPS}  &   74 & ~0.931 & ~1.660 &   --- &        &        &        \\
\citet{Schinzel2015_2FGL}        &   50 & ~0.351 & ~0.400 &   --- &        &        &        \\
NPCS, \citet{Popkov2021}         &   45 & 12.872 & 21.210 &   --- &        &        &        \\
DEVOS, \citet{Frey2008a}         &   22 & ~0.390 & ~0.700 &    16 & ~0.390 & ~0.700 & ~0.700 \\
JPL, \citet{Jacobs2020}          &   14 & ~0.180 & ~0.222 &   --- &        &        &        \\
\citet{Coppejans2016}            &   10 & ~1.008 & ~1.150 &    10 & ~1.008 & ~1.150 & ~1.150 \\
AUS CRF soluiton (2)             &    9 & ~0.611 & ~0.915 &    11 & ~1.297 & ~1.726 & ~1.977 \\
QCAL-1, \citet{Petrov2012_QCAL1} &    8 & 17.433 & 16.500 &   --- &        &        &        \\
OPA CRF soluiton (2)             &    6 & ~0.361 & ~0.501 &     1 & ~3.590 & ~3.484 & ~4.420 \\
\citet{Frey2010}                 &    5 & ~0.494 & ~0.500 &     4 & ~0.714 & ~0.750 & ~0.797 \\
\citet{Winn2002}                 &    5 & 29.587 & 60.000 &     1 & 29.587 & 60.000 & 60.000 \\
BKG CRF soluiton (2)             &    3 & ~2.836 & ~3.473 &   --- &        &        &        \\
\citet{Cao2017}                  &    3 & ~0.489 & ~0.500 &     1 & ~0.500 & ~0.500 & ~0.500 \\
\citet{Li2018}                   &    3 & ~0.611 & ~0.900 &     3 & ~0.611 & ~0.900 & ~0.900 \\
\citet{Momjian2004}              &    3 & ~1.000 & ~1.000 &     2 & ~0.755 & ~1.000 & ~1.000 \\
LCS1, \citet{Petrov2011_LCS1}    &    2 & ~2.764 & ~2.345 &   --- &        &        &        \\
\citet{Cao2014}                  &    1 & ~0.600 & ~0.600 &     1 & ~0.600 & ~0.600 & ~0.600 \\
\citet{Cao2019}                  &    1 & ~0.055 & ~0.200 &     1 & ~0.055 & ~0.200 & ~0.200 \\
\citet{Fan2020}                  &    1 & ~0.796 & ~0.600 &     1 & ~0.796 & ~0.600 & ~0.796 \\
\citet{Frey2005}                 &    1 & ~2.000 & ~2.000 &     1 & ~2.000 & ~2.000 & ~2.000 \\
\citet{Frey2008b}                &    1 & ~0.837 & ~1.000 &     1 & ~0.837 & ~1.000 & ~1.000 \\
\citet{Frey2011}                 &    1 & ~0.231 & ~0.400 &     1 & ~0.231 & ~0.400 & ~0.400 \\
\citet{Frey2013}                 &    1 & ~0.694 & ~1.000 &   --- &        &        &        \\
\citet{Gabanyi2015}              &    1 & ~0.998 & ~1.000 &     1 & ~0.998 & ~1.000 & ~1.000 \\
\citet{Gabanyi2018}              &    1 & ~0.982 & ~1.000 &     1 & ~0.982 & ~1.000 & ~1.000 \\
\citet{Gabanyi2019}              &    1 & ~2.942 & ~3.000 &     1 & ~2.942 & ~3.000 & ~3.000 \\
\citet{Gabanyi2023}              &    1 & ~0.331 & ~0.570 &     1 & ~0.331 & ~0.570 & ~0.570 \\
\citet{Marcote2018}              &    1 & ~0.722 & ~0.800 &     1 & ~0.722 & ~0.800 & ~0.800 \\
\citet{Momjian2021}              &    1 & ~9.468 & 10.000 &     1 & ~9.468 & 10.000 & 10.000 \\
OBRS-1, \citet{Petrov2011_OBRS1} &    1 & ~4.919 & ~7.170 &   --- &        &        &        \\
VIE CRF soluiton (3)             &    1 & ~1.704 & ~3.462 &   --- &        &        &        \\
\citet{Yang2012}                 &    1 & ~0.618 & ~1.000 &     1 & ~0.618 & ~1.000 & ~1.000 \\
\citet{Krezinger2024}            &  --- &        &        &     9 & ~1.264 & ~1.300 & ~1.300 \\
\citet{Gabanyi2025}              &  --- &        &        &     1 & ~0.139 & ~0.200 & ~0.200 \\
\hline
\end{tabular}
\end{center}
\textbf{Note.}
Column (1):
(1) \url{https://crf.usno.navy.mil},
(2) \url{https://cddis.nasa.gov/archive/vlbi/ivsproducts/crf},
(3) \url{https://www.vlbi.at/index.php/products}.
Columns (2,5): the number of sources used in OCARS.
\end{table*}

The source positions in the ICRF, USN and RFC catalogs were derived
using a least squares (LS) adjustment.
However, the published position uncertainties in these catalogs were
not taken directly from the LS solution, but were modified (inflated)
using the formula $\sigma_{cat} = \sqrt{(s\,\sigma_{ls})^2+\sigma_0^2}$,
where $\sigma_{cat}$ is the final (published) catalog position uncertainty,
$\sigma_{ls}$ is the formal error obtained from the LS solution,
$s$ is the scaling factor, and $\sigma_0$ is the noise floor.
In the ICRF3-SX and USN catalogs, a scaling factor of 1.5 and
an noise floor of 30~$\mu$as for both R.A. and decl. were applied
\citep{Charlot2020}.
In the RFC, a scaling factor of 1.08 for R.A. and 1.16 for decl.,
and a declination-dependent noise floor in the range 25--132~$\mu$as for R.A.
and 92--215~$\mu$as for decl. were used \citep{Petrov2025_RFC}.               %%%% Appendix, Table 12
In old OCARS catalog, unmodified USN position uncertainties were used
(the published uncertainties of the source coordinates were deflated
and descaled using the above-mentioned parameters).
Effective 2025, the inflated USN position uncertainties are used in OCARS.
This explains the large difference in USN data in Table~\ref{tab:ocars_catalogs}
for the old and new OCARS releases.

In addition to the sources taken from the catalogs and experiments
listed above, several tens of prospective sources from \citet{Bourda2010},
databases stored in the IVS data center, and Solve \texttt{blokq}
file\footnote{\url{https://cddis.nasa.gov/archive/vlbi/gsfc/ancillary/solve_apriori/blokq.c11.dat}}
are also included in OCARS.
Typically, these sources have been attempted to be observed with astrometric VLBI,
but without success (not detected, not enough good observations to obtain reliable
coordinates, etc.).
The positions of these sources are mostly preliminary, but they are stored
in the OCARS source list to encourage their repeated observations.
Besides, a few objects in OCARS correspond to the cores of multi-component
sources and lensed images, which helps the user more easily identify ambiguities
in observed source positions.

Until the end of 2024, source coordinates were simply copied from different
journal and online publications.
When a source was present in several catalogs, the coordinates were taken from them
in the following order of preference.
The first releases of OCARS used the ICRF \citep{Ma1998} catalog as the primary
source of coordinates, which was replaced by ICRF2 \citep{Fey2015} in June 2010,
which in turn was replaced by ICRF3 \citep{Charlot2020} in July 2018.
The latter is comprised from three catalogs obtained from observations
in different wavebands: ICRF3-SX (4536 sources), ICRF-K (824 sources),
and ICRF-XKa (678 sources).
The coordinates of sources present in two or three ICRF3 catalogs were
included in OCARS in the same order.

As the number of astrometric radio sources and the number of their observations
increased over time, the ICRF3 catalogs became outdated,
and since March 2023, the USN catalogs have been used first, and then
other catalogs from the IVS Data Center if they contain sources not
included in the USN catalogs.
The USN catalogs were chosen as the primary ones because they
are computed using the same software and strategy used for computation
of the ICRF3-SX catalog, which is supposed to be the primary
implementation of the ICRS in radio.
Sources not in the IVS catalogs have been taken from other publications,
usually in the order of their publication.

Up to 2025, a total of 40 catalogs and special experiments were used
to compile the OCARS source list, see the left side of Table~\ref{tab:ocars_catalogs}.
All input catalogs, except USN, do not have global sky coverage
and/or contain too few sources, which makes it practically impossible
to transform them to a unified coordinate frame.
As a consequence, the source positions in OCARS were systematically inhomogeneous.

The situation changed significantly with the publication of the Radio
Fundamental Catalog (RFC) \citep{Petrov2025_RFC}.
It accumulates about a dozen smaller catalogs previously used in OCARS,
includes a large number of previously unpublished VLBI source positions,
and has global sky coverage.
This makes it possible to reliably assess the orientation of the catalog
and higher-order systematics.
In the latest OCARS release (March 2025) the number of input catalogs
and experiments has been reduced to 28
(including two new ones that were not used in the previous OCARS release),
with 99.3\% of the 22,087 OCARS sources concentrated in two of them,
USN (version 2024b, 5825 sources) and
RFC (version 2024c, 16108 sources used in OCARS), see the right side
of Table~\ref{tab:ocars_catalogs}.

Using this capability, improved OCARS positions were computed starting
from March 2025.
For this purpose, the differences in the coordinates of sources
in the USN and RFC catalogs with those in the ICRF3-SX catalog
were modeled using the vector spherical harmonics (VSH) decomposition.
While most recent VLBI-based radio source catalog position comparisons
were performed using the 16-term VSH decomposition to degree 2
(see Appendix for details), for this study I used the 48-term
decomposition to degree 4, following \citet{Makarov2023}.
Explicit expressions for the VSH functions for this decomposition
are given in the Appendix.

All USN and RFC sources, not only ICRF3 defining ones, were used
in assessing their systematics with respect to the ICRF3-SX.
Outliers were detected and rejected using the following criteria.
A source is considered an outlier if the angular separation between
the source positions in the two catalogs is greater than 5~mas, or
the error in the position difference is greater than 5~mas, or
the normalized separation between the source positions in the two
catalogs (as defined in \citet{Mignard2016}) is greater than 5.
As for the difference between the USN and ICRF3-SX catalogs,
they have a total of 4534 common sources (two ICRF3-SX sources
were not included in the USN catalog release used in this work),
of which 129 sources were rejected, resulting in 4405 common sources.
As for the difference between the RFC and ICRF3-SX catalogs,
they have a total of 4536 common sources, of which 134 sources
were rejected, resulting in 4402 common sources.

The authors of the RFC catalog used all available astrometric
and geodetic VLBI data, independently on the wavelength at which
the observations were performed.
Therefore, this dataset includes, presumably, all the S/X databases
that were used to derive the USN catalog.
As a consequence, all the ICRF sources are present in the RFC catalog,
and the list of sources in common between these two catalogs is,
in fact, the ICRF-SX source list.

\begin{figure*}
\begin{center}
\includegraphics[clip,width=\textwidth]{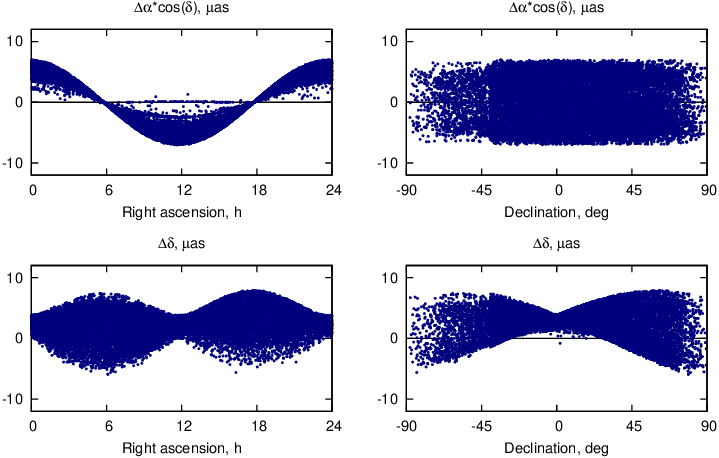}
\caption{Differences between the RFC catalog transformed to the ICRF3
  GA model and the original RFC catalog.}
\label{fig:rfcga-rfc}
\end{center}
\end{figure*}

Before comparison, USN and RFC catalogs should be unified with
the ICRF3 in modeling the Galactic aberration (GA)
\citep{Kovalevsky2003,Kopeikin2006,Titov2013,Malkin2014,MacMillan2019}.
Two main parameters are used in modeling the GA effect:
the GA constant $A$ and the reference epoch $t_0$.
For both ICRF3 and USN catalogs (as well as for other catalogs computed
in IVS analysis centers), the same parameters $A=\;$\muasy{5.8}
and $t_0=2015.0$ are used, while the RFC catalog was computed with
different parameters $A=\;$\muasy{5.49} and $t_0=2016.0$.
Therefore, the RFC was converted to the same GA parameters used in ICRF3.
Accordingly, the OCARS position epoch for RFC sources was set to 2015.0.
It should be noted that the effect of the difference in the GA constant
(5.8 in ICRF3 and USN versus 5.49 in RFC) for a given source depends on
the time interval between the average observation epoch for this source
and the reference GA epoch.
The differences between the transformed and original RFC catalogs are shown
in Fig.~\ref{fig:rfcga-rfc}.
They reach \muas{7.9}, which is significant at the level
of accuracy of modern astrometric catalogs.

The GA effect causes a change in the glide coefficients
$\ve{S}_{10}$ (\muas{2.8} in the differences between the transformed and original
RFC catalogs), $\ve{S}_{11}^{\rm Re}$ (\muas{0.3}),
and $\ve{S}_{11}^{\rm Im}$ (\muas{5.0})
of the VSH decomposition \citep{Titov2011,Klioner2021} (see Appendix)
which can be considered as a compensation of the effect of different GA modeling
in the ICRF and RFC catalogs.
However, such an approach makes the position epoch in transformed catalog uncertain.
Therefore, it is preferable to transform the input catalogs to the same GA model
prior applying the VSH analysis.

This procedure does not take into account Galactic sources with large proper motions.
For example, in the RFC catalog, 25 sources were classified by the authors
as Galactic objects\footnote{This classification not always coincides with OCARS,
but analysis of these discrepancies is beyond the scope of this paper.},
and statistically significant proper motions were found for 17 of them.
However, these sources are rarely included in the radio astrometry studies,
are provided with relevant comments in OCARS, and can be analysed in detail
separately by interested user.

\begin{figure*}
\begin{center}
\includegraphics[clip,width=\textwidth]{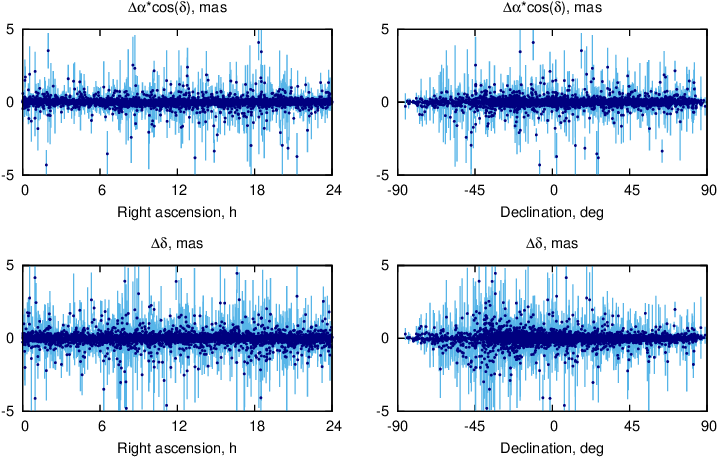}\\[3em]
\includegraphics[clip,width=\textwidth]{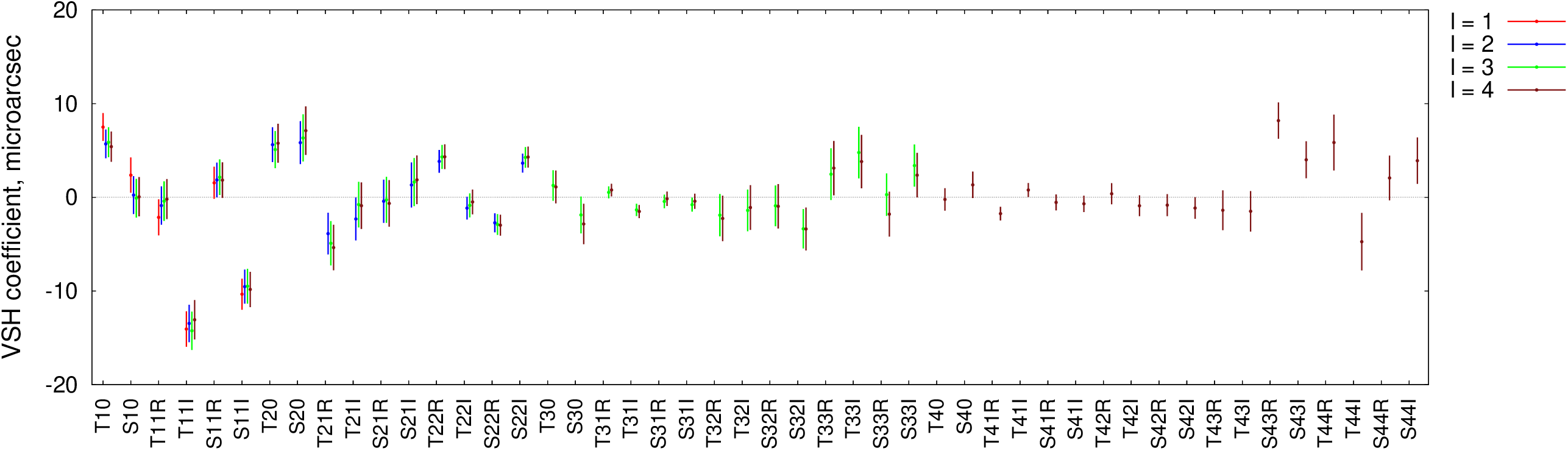}
\caption{Differences between the USN and ICRF3-SX catalogs (USN minus ICRF3-SX)
  after removing outliers.
  The bottom panel shows results of VSH decomposition for degrees $l=1 \ldots 4$.}
\label{fig:usn-icrf}
\end{center}
\end{figure*}

\begin{figure*}
\begin{center}
\includegraphics[clip,width=\textwidth]{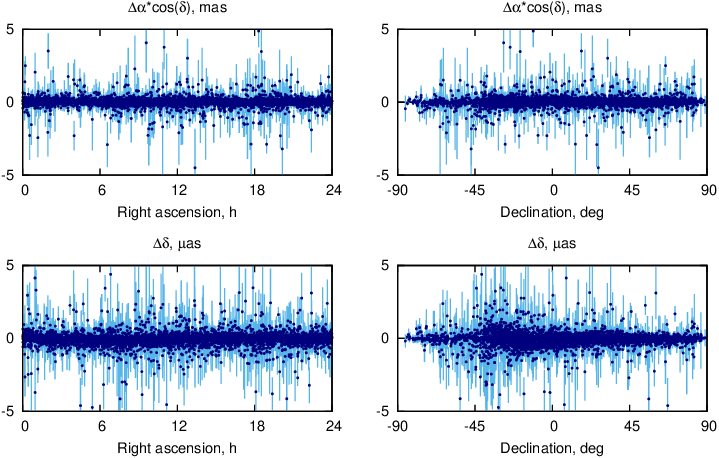}\\[3em]
\includegraphics[clip,width=\textwidth]{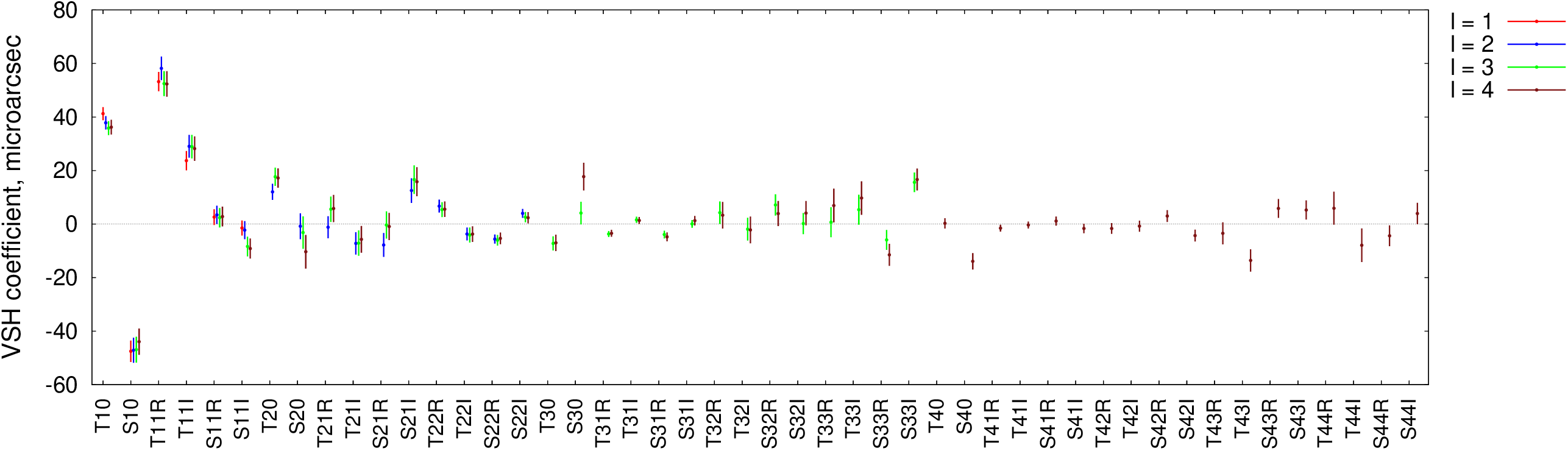}
\caption{Differences between the RFC and ICRF3-SX catalogs (RFC minus ICRF3-SX)
  after removing outliers
  The bottom panel shows results of VSH decomposition for degrees $l=1 \ldots 4$.}
\label{fig:rfc-icrf}
\end{center}
\end{figure*}

Results of the comparison of the USN and RFC with ICRF3-SX are presented
in Figs.~\ref{fig:usn-icrf} and \ref{fig:rfc-icrf}.
It should be noted that although the differences between the
source coordinates in two catalogs approximated by VSH (or other functions)
are traditionally called ``systematic'', they actually include both systematic
and random position errors and can be considered more strictly as large-scale
sky-correlated errors as discussed in detail by \citet{Makarov2012}.

This analysis shows that the systematic differences between the USN and
ICRF3-SX catalogs are much smaller than for the RFC.
There may be several reasons for that.
The two main factors that cause these discrepancies are most likely the following.
The USN catalog is more systematically homogeneous since it is based only
on the processing of observations made in the S/X wavebands, while the RFC
catalog is computed using observations made at different wavelengths on both
dual and single frequencies.
Another difference between the USN and RFC analyses is that the USN catalog
was aligned to ICRF3, while the RFC was aligned to the outdated ICRF catalog.

For the USN catalog, all the VSH coefficients are below 15$\mu$as,
which in turn is below the ICRF axes stability level previously estimated
as $\sim$\muas{20} \citep{Lambert2013,LiuN2022,Malkin2024_ICRS}.
Nevertheless, this rotation was applied to the USN catalog,
for the rigor of the final solution.

For a quantitative assessment of the improvement in position
differences between the USN and RFC catalogs with the ICRF3-SX catalogs,
the weighted root-mean-square (WRMS) differences between the original
and converted USN/RFC and the ICRF3-SX catalogs were computed.
For the USN catalog, the WRMS(original) are 0.0844~mas and 0.0967~mas
for R.A. and decl., respectively,
and the WRMS(converted) are 0.0823~mas and 0.0958~mas.
For the RFC, the WRMS(original) are 0.1433~mas and 0.2083~mas,
and the WRMS(converted) are 0.1349~mas and 0.1960~mas.
Comparison of WRMS before and after conversion to the ICRF3-SX
system using an F-test applied to the variance ratio
\begin{equation*}
\mathrm{WRMS}^2\mathrm{(original)} \ / \ \mathrm{WRMS}^2\mathrm{(converted)}
\end{equation*}
showed a moderate improvement for USN sources with a probability
of 94.4\% for R.A. and 72.3\% for decl., while for RFC sources,
a significant improvement was observed for both coordinates with
a probability greater than 99.99\%.

\begin{figure*}
\begin{center}
\includegraphics[clip,width=\textwidth]{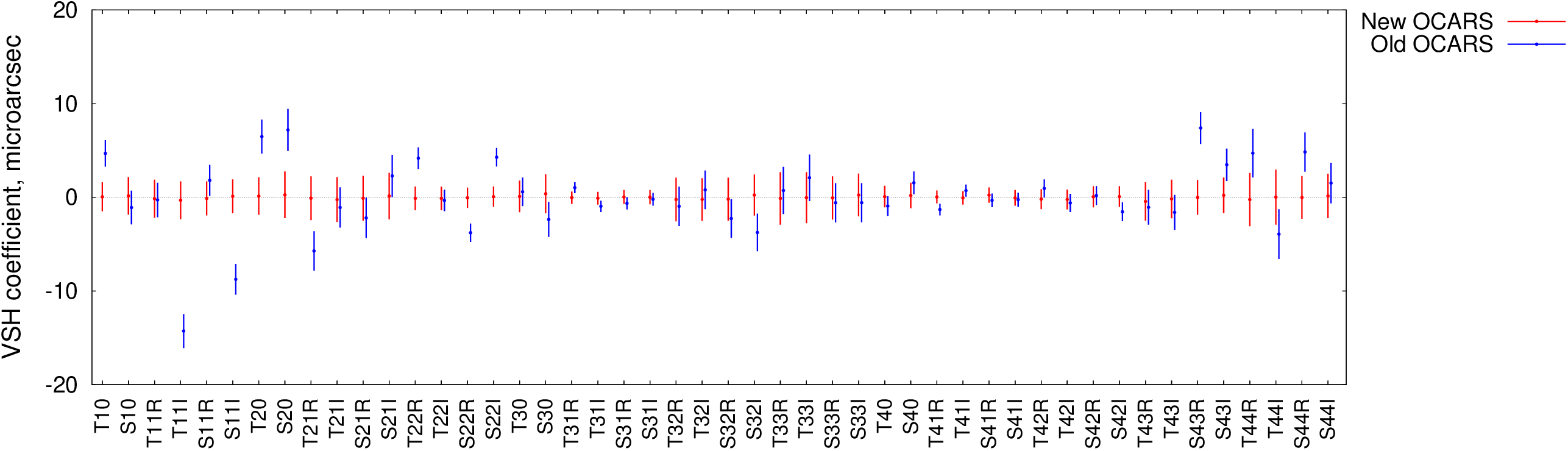}
\caption{VSH decomposition of the differences between the OCARS and ICRF3-SX catalogs (OCARS minus ICRF3-SX).}
\label{fig:ocars-icrf-vsh}
\end{center}
\end{figure*}

The main result of this work is presented in Fig.~\ref{fig:ocars-icrf-vsh},
which shows how much the systematic differences between the OCARS and ICRF3
catalogs have improved after the implementation of the procedures described above.

%%%%%%%%%%%%%%%%%%%%%%%%%%%%%%%%%%%%%%%%%%%%%%%%%%%%%%%%%%%%%%%%%%%%%%%%%%%%%%%%%%%%%%%%%%%%%%%%%%%%%%%%%%%%%%

\section{Conclusion}
\label{sect:conclusions}

OCARS is a compiled catalog created and maintained with the primary goal of collecting
in one place and conveniently providing coordinates and physical properties
of astrometric and geodetic radio sources collected from the literature and other
data sources.
created in 2007 to 22094 in March 2025.
The number of sources in OCARS has grown from 3914 in the first version created
in 2007 to 22094 in March 2025.
In the latest OCARS release from March 2025, redshift information is available
for 62.3\% of sources (80\% of them are spectroscopic), and optical and NIR photometry
is available for 85.3\% of sources.
For ICRF3 sources, redshift information is available for 80.2\% of sources
(89.9\% of them are spectroscopic), and optical and NIR photometry is available
for 94.7\% of sources.
The catalog is updated regularly, on average every few weeks, following the publication
of new radio radio astrometric catalogs or as new information appears in databases and publications.
A total of 179 OCARS releases have been issued to date.

In this work, a first attempt was made to create a systematically homogeneous
set of coordinates of all OCARS sources in the ICRF3 system.
The primary initial contribution to new OCARS positions comes from the RFC catalog
(about 16,000 sources) and the USNO catalog (about 6,000 sources),
which together account for 99.3\% of all OCARS sources.
Both of these catalogs were transformed into the ICRF3-SX system using
the 48-term VSH decomposition.
It turned out that the transformation parameters (VSH coefficients) for
the RFC catalog are several times greater than the transformation parameters
for the USNO catalog, especially for VSH of degree 1.

Comparison of the source coordinates in the new OCARS version with the ICRF3-SX
catalog showed practically complete elimination of systematic errors in OCARS
positions relative to the ICRF3-SX catalog.
With this new set of radio source coordinates, the OCARS catalog can be considered
for some purposes as an extension of the ICRF3 system to 20+ thousand sources,
which may allow for a significant increase in the reliability of comparisons
between radio (ICRF) and optical ($Gaia$-CRF) ICRS realizations.

%%%%%%%%%%%%%%%%%%%%%%%%%%%%%%%%%%%%%%%%%%%%%%%%%%%%%%%%%%%%%%%%%%%%%%%%%%%%%%%%%%%%%%%%%%%%%%%%%%%%%%%%%%%%%%

\section*{Acknowledgments}

The author expresses deep gratitude to his colleagues who made the results
of their determinations of the radio source positions available to the community.
The author thanks the anonymous reviewer for useful comments and suggestions
which helped to improve the manuscript.
This research has made use of the SAO/NASA Astrophysics Data
System (ADS), \url{https://ui.adsabs.harvard.edu/}.
Figures were prepared using the \texttt{gnuplot}, \url{http://www.gnuplot.info/}.
Computations related to the $F$-test statistics were performed using the
\texttt{Critical Value Calculator}, \url{https://statcalculator.net/critical-value-calculator}.

%%%%%%%%%%%%%%%%%%%%%%%%%%%%%%%%%%%%%%%%%%%%%%%%%%%%%%%%%%%%%%%%%

\bibliography{ocars_positions.bib}

\begin{thebibliography}{}
\expandafter\ifx\csname natexlab\endcsname\relax\def\natexlab#1{#1}\fi
\providecommand{\url}[1]{\href{#1}{#1}}
\providecommand{\dodoi}[1]{doi:~\href{http://doi.org/#1}{\nolinkurl{#1}}}
\providecommand{\doeprint}[1]{\href{http://ascl.net/#1}{\nolinkurl{http://ascl.net/#1}}}
\providecommand{\doarXiv}[1]{\href{https://arxiv.org/abs/#1}{\nolinkurl{https://arxiv.org/abs/#1}}}

\bibitem[{{Bourda} {et~al.}(2010){Bourda}, {Charlot}, {Porcas}, \&
  {Garrington}}]{Bourda2010}
{Bourda}, G., {Charlot}, P., {Porcas}, R.~W., \& {Garrington}, S.~T. 2010,
  \aap, 520, A113, \dodoi{10.1051/0004-6361/201014248}

\bibitem[{{Cao} {et~al.}(2017){Cao}, {Frey}, {Gab{\'a}nyi}, {Paragi}, {Yang},
  {Cseh}, {Hong}, \& {An}}]{Cao2017}
{Cao}, H.~M., {Frey}, S., {Gab{\'a}nyi}, K.~{\'E}., {et~al.} 2017, \mnras, 467,
  950, \dodoi{10.1093/mnras/stx160}

\bibitem[{{Cao} {et~al.}(2019){Cao}, {Frey}, {Gab{\'a}nyi}, {Yang}, {Cui},
  {Hong}, \& {An}}]{Cao2019}
{Cao}, H.-M., {Frey}, S., {Gab{\'a}nyi}, K.~{\'E}., {et~al.} 2019, \mnras, 482,
  L34, \dodoi{10.1093/mnrasl/sly184}

\bibitem[{{Cao} {et~al.}(2014){Cao}, {Frey}, {Gurvits}, {Yang}, {Hong},
  {Paragi}, {Deller}, \& {Ivezi{\'c}}}]{Cao2014}
{Cao}, H.~M., {Frey}, S., {Gurvits}, L.~I., {et~al.} 2014, \aap, 563, A111,
  \dodoi{10.1051/0004-6361/201323328}

\bibitem[{{Charlot} {et~al.}(2020){Charlot}, {Jacobs}, {Gordon}, {Lambert}, {de
  Witt}, {B{\"o}hm}, {Fey}, {Heinkelmann}, {Skurikhina}, {Titov}, {Arias},
  {Bolotin}, {Bourda}, {Ma}, {Malkin}, {Nothnagel}, {Mayer}, {MacMillan},
  {Nilsson}, \& {Gaume}}]{Charlot2020}
{Charlot}, P., {Jacobs}, C.~S., {Gordon}, D., {et~al.} 2020, \aap, 644, A159,
  \dodoi{10.1051/0004-6361/202038368}

\bibitem[{{Coppejans} {et~al.}(2016){Coppejans}, {Frey}, {Cseh}, {M{\"u}ller},
  {Paragi}, {Falcke}, {Gab{\'a}nyi}, {Gurvits}, {An}, \&
  {Titov}}]{Coppejans2016}
{Coppejans}, R., {Frey}, S., {Cseh}, D., {et~al.} 2016, \mnras, 463, 3260,
  \dodoi{10.1093/mnras/stw2236}

\bibitem[{{Fan} {et~al.}(2020){Fan}, {Chen}, {An}, {Xie}, {Han}, {Knudsen}, \&
  {Yang}}]{Fan2020}
{Fan}, L., {Chen}, W., {An}, T., {et~al.} 2020, \apjl, 905, L32,
  \dodoi{10.3847/2041-8213/abcebf}

\bibitem[{{Fey} {et~al.}(2015){Fey}, {Gordon}, {Jacobs}, {Ma}, {Gaume},
  {Arias}, {Bianco}, {Boboltz}, {B{\"o}ckmann}, {Bolotin}, {Charlot},
  {Collioud}, {Engelhardt}, {Gipson}, {Gontier}, {Heinkelmann}, {Kurdubov},
  {Lambert}, {Lytvyn}, {MacMillan}, {Malkin}, {Nothnagel}, {Ojha},
  {Skurikhina}, {Sokolova}, {Souchay}, {Sovers}, {Tesmer}, {Titov}, {Wang}, \&
  {Zharov}}]{Fey2015}
{Fey}, A.~L., {Gordon}, D., {Jacobs}, C.~S., {et~al.} 2015, \aj, 150, 58,
  \dodoi{10.1088/0004-6256/150/2/58}

\bibitem[{{Frey} {et~al.}(2008{\natexlab{a}}){Frey}, {Gurvits}, {Paragi}, \&
  {{\'E}. Gab{\'a}nyi}}]{Frey2008b}
{Frey}, S., {Gurvits}, L.~I., {Paragi}, Z., \& {{\'E}. Gab{\'a}nyi}, K.
  2008{\natexlab{a}}, \aap, 484, L39, \dodoi{10.1051/0004-6361:200810040}

\bibitem[{{Frey} {et~al.}(2008{\natexlab{b}}){Frey}, {Gurvits}, {Paragi},
  {Mosoni}, {Garrett}, \& {Garrington}}]{Frey2008a}
{Frey}, S., {Gurvits}, L.~I., {Paragi}, Z., {et~al.} 2008{\natexlab{b}}, \aap,
  477, 781, \dodoi{10.1051/0004-6361:20078711}

\bibitem[{{Frey} {et~al.}(2013){Frey}, {Paragi}, {Gab{\'a}nyi}, \&
  {An}}]{Frey2013}
{Frey}, S., {Paragi}, Z., {Gab{\'a}nyi}, K.~{\'E}., \& {An}, T. 2013, \aap,
  552, A109, \dodoi{10.1051/0004-6361/201220778}

\bibitem[{{Frey} {et~al.}(2010){Frey}, {Paragi}, {Gurvits}, {Cseh}, \&
  {Gab{\'a}nyi}}]{Frey2010}
{Frey}, S., {Paragi}, Z., {Gurvits}, L.~I., {Cseh}, D., \& {Gab{\'a}nyi},
  K.~{\'E}. 2010, \aap, 524, A83, \dodoi{10.1051/0004-6361/201015554}

\bibitem[{{Frey} {et~al.}(2011){Frey}, {Paragi}, {Gurvits}, {Gab{\'a}nyi}, \&
  {Cseh}}]{Frey2011}
{Frey}, S., {Paragi}, Z., {Gurvits}, L.~I., {Gab{\'a}nyi}, K.~{\'E}., \&
  {Cseh}, D. 2011, \aap, 531, L5, \dodoi{10.1051/0004-6361/201117341}

\bibitem[{{Frey} {et~al.}(2005){Frey}, {Paragi}, {Mosoni}, \&
  {Gurvits}}]{Frey2005}
{Frey}, S., {Paragi}, Z., {Mosoni}, L., \& {Gurvits}, L.~I. 2005, \aap, 436,
  L13, \dodoi{10.1051/0004-6361:200500112}

\bibitem[{{Gab{\'a}nyi} {et~al.}(2015){Gab{\'a}nyi}, {Cseh}, {Frey}, {Paragi},
  {Gurvits}, {An}, \& {Zhang}}]{Gabanyi2015}
{Gab{\'a}nyi}, K.~E., {Cseh}, D., {Frey}, S., {et~al.} 2015, \mnras, 450, L57,
  \dodoi{10.1093/mnrasl/slv046}

\bibitem[{{Gab{\'a}nyi} {et~al.}(2018){Gab{\'a}nyi}, {Frey}, {Gurvits},
  {Paragi}, \& {Perger}}]{Gabanyi2018}
{Gab{\'a}nyi}, K.~{\'E}., {Frey}, S., {Gurvits}, L.~I., {Paragi}, Z., \&
  {Perger}, K. 2018, Research Notes of the American Astronomical Society, 2,
  200, \dodoi{10.3847/2515-5172/aaec82}

\bibitem[{{Gab{\'a}nyi} {et~al.}(2025){Gab{\'a}nyi}, {Frey}, {Perger}, \&
  {Kun}}]{Gabanyi2025}
{Gab{\'a}nyi}, K.~{\'E}., {Frey}, S., {Perger}, K., \& {Kun}, E. 2025,
  Universe, 11, 83, \dodoi{10.3390/universe11030083}

\bibitem[{{Gab{\'a}nyi} {et~al.}(2019){Gab{\'a}nyi}, {Frey}, {Satyapal},
  {Constantin}, \& {Pfeifle}}]{Gabanyi2019}
{Gab{\'a}nyi}, K.~{\'E}., {Frey}, S., {Satyapal}, S., {Constantin}, A., \&
  {Pfeifle}, R.~W. 2019, \aap, 630, L5, \dodoi{10.1051/0004-6361/201936519}

\bibitem[{{Gab{\'a}nyi} {et~al.}(2023){Gab{\'a}nyi}, {Belladitta}, {Frey},
  {Orosz}, {Gurvits}, {Rozgonyi}, {An}, {Cao}, {Paragi}, \&
  {Perger}}]{Gabanyi2023}
{Gab{\'a}nyi}, K.~{\'E}., {Belladitta}, S., {Frey}, S., {et~al.} 2023, \pasa,
  40, e004, \dodoi{10.1017/pasa.2023.2}

\bibitem[{{Gaia Collaboration} {et~al.}(2021){Gaia Collaboration}, {Klioner},
  {Mignard}, {Lindegren}, {Bastian}, {McMillan}, {Hern{\'a}ndez}, {Hobbs},
  {Ramos-Lerate}, {Biermann}, {Bombrun}, {de Torres}, {Gerlach}, {Geyer},
  {Hilger}, {Lammers}, {Steidelm{\"u}ller}, {Stephenson}, {Brown}, {Vallenari},
  {Prusti}, {de Bruijne}, {Babusiaux}, {Creevey}, {Evans}, {Eyer}, {Hutton},
  {Jansen}, {Jordi}, {Luri}, {Panem}, {Pourbaix}, {Randich}, {Sartoretti},
  {Soubiran}, {Walton}, {Arenou}, {Bailer-Jones}, {Cropper}, {Drimmel}, {Katz},
  {Lattanzi}, {van Leeuwen}, {Bakker}, {Casta{\~n}eda}, {De Angeli},
  {Ducourant}, {Fabricius}, {Fouesneau}, {Fr{\'e}mat}, {Guerra}, {Guerrier},
  {Guiraud}, {Jean-Antoine Piccolo}, {Masana}, {Messineo}, {Mowlavi},
  {Nicolas}, {Nienartowicz}, {Pailler}, {Panuzzo}, {Riclet}, {Roux},
  {Seabroke}, {Sordo}, {Tanga}, {Th{\'e}venin}, {Gracia-Abril}, {Portell},
  {Teyssier}, {Altmann}, {Andrae}, {Bellas-Velidis}, {Benson}, {Berthier},
  {Blomme}, {Brugaletta}, {Burgess}, {Busso}, {Carry}, {Cellino}, {Cheek},
  {Clementini}, {Damerdji}, {Davidson}, {Delchambre}, {Dell'Oro},
  {Fern{\'a}ndez-Hern{\'a}ndez}, {Galluccio}, {Garc{\'\i}a-Lario},
  {Garcia-Reinaldos}, {Gonz{\'a}lez-N{\'u}{\~n}ez}, {Gosset}, {Haigron},
  {Halbwachs}, {Hambly}, {Harrison}, {Hatzidimitriou}, {Heiter}, {Hestroffer},
  {Hodgkin}, {Holl}, {Jan{\ss}en}, {Jevardat de Fombelle}, {Jordan},
  {Krone-Martins}, {Lanzafame}, {L{\"o}ffler}, {Lorca}, {Manteiga}, {Marchal},
  {Marrese}, {Moitinho}, {Mora}, {Muinonen}, {Osborne}, {Pancino}, {Pauwels},
  {Recio-Blanco}, {Richards}, {Riello}, {Rimoldini}, {Robin}, {Roegiers},
  {Rybizki}, {Sarro}, {Siopis}, {Smith}, {Sozzetti}, {Ulla}, {Utrilla}, {van
  Leeuwen}, {van Reeven}, {Abbas}, {Abreu Aramburu}, {Accart}, {Aerts},
  {Aguado}, {Ajaj}, {Altavilla}, {{\'A}lvarez}, {{\'A}lvarez Cid-Fuentes},
  {Alves}, {Anderson}, {Anglada Varela}, {Antoja}, {Audard}, {Baines}, {Baker},
  {Balaguer-N{\'u}{\~n}ez}, {Balbinot}, {Balog}, {Barache}, {Barbato},
  {Barros}, {Barstow}, {Bartolom{\'e}}, {Bassilana}, {Bauchet},
  {Baudesson-Stella}, {Becciani}, {Bellazzini}, {Bernet}, {Bertone}, {Bianchi},
  {Blanco-Cuaresma}, {Boch}, {Bossini}, {Bouquillon}, {Bramante}, {Breedt},
  {Bressan}, {Brouillet}, {Bucciarelli}, {Burlacu}, {Busonero}, {Butkevich},
  {Buzzi}, {Caffau}, {Cancelliere}, {C{\'a}novas}, {Cantat-Gaudin}, {Carballo},
  {Carlucci}, {Carnerero}, {Carrasco}, {Casamiquela}, {Castellani},
  {Castro-Ginard}, {Castro Sampol}, {Chaoul}, {Charlot}, {Chemin}, {Chiavassa},
  {Comoretto}, {Cooper}, {Cornez}, {Cowell}, {Crifo}, {Crosta}, {Crowley},
  {Dafonte}, {Dapergolas}, {David}, {David}, {de Laverny}, {De Luise}, {De
  March}, {De Ridder}, {de Souza}, {de Teodoro}, {del Peloso}, {del Pozo},
  {Delgado}, {Delgado}, {Delisle}, {Di Matteo}, {Diakite}, {Diener},
  {Distefano}, {Dolding}, {Eappachen}, {Enke}, {Esquej}, {Fabre}, {Fabrizio},
  {Faigler}, {Fedorets}, {Fernique}, {Fienga}, {Figueras}, {Fouron},
  {Fragkoudi}, {Fraile}, {Franke}, {Gai}, {Garabato}, {Garcia-Gutierrez},
  {Garc{\'\i}a-Torres}, {Garofalo}, {Gavras}, {Giacobbe}, {Gilmore}, {Girona},
  {Giuffrida}, {Gomez}, {Gonzalez-Santamaria}, {Gonz{\'a}lez-Vidal}, {Granvik},
  {Guti{\'e}rrez-S{\'a}nchez}, {Guy}, {Hauser}, {Haywood}, {Helmi}, {Hidalgo},
  {H{\l}adczuk}, {Holland}, {Huckle}, {Jasniewicz}, {Jonker}, {Juaristi
  Campillo}, {Julbe}, {Karbevska}, {Kervella}, {Khanna}, {Kochoska},
  {Kordopatis}, {Korn}, {Kostrzewa-Rutkowska}, {Kruszy{\'n}ska}, {Lambert},
  {Lanza}, {Lasne}, {Le Campion}, {Le Fustec}, {Lebreton}, {Lebzelter},
  {Leccia}, {Leclerc}, {Lecoeur-Taibi}, {Liao}, {Licata}, {Lindstr{\o}m},
  {Lister}, {Livanou}, {Lobel}, {Madrero Pardo}, {Managau}, {Mann}, {Marchant},
  {Marconi}, {Marcos Santos}, {Marinoni}, {Marocco}, {Marshall}, {Martin Polo},
  {Mart{\'\i}n-Fleitas}, {Masip}, {Massari}, {Mastrobuono-Battisti}, {Mazeh},
  {Messina}, {Michalik}, {Millar}, {Mints}, {Molina}, {Molinaro}, {Moln{\'a}r},
  {Montegriffo}, {Mor}, {Morbidelli}, {Morel}, {Morris}, {Mulone}, {Munoz},
  {Muraveva}, {Murphy}, {Musella}, {Noval}, {Ord{\'e}novic}, {Orr{\`u}},
  {Osinde}, {Pagani}, {Pagano}, {Palaversa}, {Palicio}, {Panahi}, {Pawlak},
  {Pe{\~n}alosa Esteller}, {Penttil{\"a}}, {Piersimoni}, {Pineau}, {Plachy},
  {Plum}, {Poggio}, {Poretti}, {Poujoulet}, {Pr{\v{s}}a}, {Pulone}, {Racero},
  {Ragaini}, {Rainer}, {Raiteri}, {Rambaux}, {Ramos}, {Re Fiorentin}, {Regibo},
  {Reyl{\'e}}, {Ripepi}, {Riva}, {Rixon}, {Robichon}, {Robin}, {Roelens},
  {Rohrbasser}, {Romero-G{\'o}mez}, {Rowell}, {Royer}, {Rybicki}, {Sadowski},
  {Sagrist{\`a} Sell{\'e}s}, {Sahlmann}, {Salgado}, {Salguero}, {Samaras},
  {Sanchez Gimenez}, {Sanna}, {Santove{\~n}a}, {Sarasso}, {Schultheis},
  {Sciacca}, {Segol}, {Segovia}, {S{\'e}gransan}, {Semeux}, {Siddiqui},
  {Siebert}, {Siltala}, {Slezak}, {Smart}, {Solano}, {Solitro}, {Souami},
  {Souchay}, {Spagna}, {Spoto}, {Steele}, {S{\"u}veges}, {Szabados},
  {Szegedi-Elek}, {Taris}, {Tauran}, {Taylor}, {Teixeira}, {Thuillot},
  {Tonello}, {Torra}, {Torra}, {Turon}, {Unger}, {Vaillant}, {van Dillen},
  {Vanel}, {Vecchiato}, {Viala}, {Vicente}, {Voutsinas}, {Weiler}, {Wevers},
  {Wyrzykowski}, {Yoldas}, {Yvard}, {Zhao}, {Zorec}, {Zucker}, {Zurbach}, \&
  {Zwitter}}]{Klioner2021}
{Gaia Collaboration}, {Klioner}, S.~A., {Mignard}, F., {et~al.} 2021, \aap,
  649, A9, \dodoi{10.1051/0004-6361/202039734}

\bibitem[{{Immer} {et~al.}(2011){Immer}, {Brunthaler}, {Reid}, {Bartkiewicz},
  {Choi}, {Menten}, {Moscadelli}, {Sanna}, {Wu}, {Xu}, {Zhang}, \&
  {Zheng}}]{Immer2011_BeSSeL}
{Immer}, K., {Brunthaler}, A., {Reid}, M.~J., {et~al.} 2011, \apjs, 194, 25,
  \dodoi{10.1088/0067-0049/194/2/25}

\bibitem[{{Jacobs} {et~al.}(2020){Jacobs}, {Kroger}, \& {Volk}}]{Jacobs2020}
{Jacobs}, C.~S., {Kroger}, P., \& {Volk}, C. 2020, {Ka-Band Radio Source
  Catalog}, Tech. Rep. DSN 810-005, 108, {JPL, CalTech, CA, USA}

\bibitem[{{Karbon} \& {Nothnagel}(2019)}]{Karbon2019}
{Karbon}, M., \& {Nothnagel}, A. 2019, \aap, 630, A101,
  \dodoi{10.1051/0004-6361/201936083}

\bibitem[{{Kopeikin} \& {Makarov}(2006)}]{Kopeikin2006}
{Kopeikin}, S.~M., \& {Makarov}, V.~V. 2006, \aj, 131, 1471,
  \dodoi{10.1086/500170}

\bibitem[{{Kovalevsky}(2003)}]{Kovalevsky2003}
{Kovalevsky}, J. 2003, \aap, 404, 743, \dodoi{10.1051/0004-6361:20030560}

\bibitem[{{Kr{\'a}sn{\'a}} {et~al.}(2023){Kr{\'a}sn{\'a}}, {Baldreich},
  {B{\"o}hm}, {B{\"o}hm}, {Gruber}, {Hellerschmied}, {Jaron}, {Kern}, {Mayer},
  {Nothnagel}, {Panzenb{\"o}ck}, \& {Wolf}}]{Krasna2023}
{Kr{\'a}sn{\'a}}, H., {Baldreich}, L., {B{\"o}hm}, J., {et~al.} 2023, \aap,
  679, A53, \dodoi{10.1051/0004-6361/202245434}

\bibitem[{{Krezinger} {et~al.}(2024){Krezinger}, {Baldini}, {Giroletti},
  {Sbarrato}, {Ghisellini}, {Giovannini}, {An}, {Gab{\'a}nyi}, \&
  {Frey}}]{Krezinger2024}
{Krezinger}, M., {Baldini}, G., {Giroletti}, M., {et~al.} 2024, \aap, 690,
  A321, \dodoi{10.1051/0004-6361/202451025}

\bibitem[{{Lambert}(2013)}]{Lambert2013}
{Lambert}, S. 2013, \aap, 553, A122, \dodoi{10.1051/0004-6361/201321320}

\bibitem[{{Li} {et~al.}(2018){Li}, {Yang}, {An}, {Paragi}, {Deller},
  {Reynolds}, {Hong}, {Wang}, {Ding}, {Xia}, {Yan}, \& {Guo}}]{Li2018}
{Li}, Z., {Yang}, J., {An}, T., {et~al.} 2018, \mnras, 476, 399,
  \dodoi{10.1093/mnras/sty210}

\bibitem[{{Liu} {et~al.}(2022){Liu}, {Lambert}, {Arias}, {Liu}, \&
  {Zhu}}]{LiuN2022}
{Liu}, N., {Lambert}, S.~B., {Arias}, E.~F., {Liu}, J.~C., \& {Zhu}, Z. 2022,
  \aap, 659, A75, \dodoi{10.1051/0004-6361/202142632}

\bibitem[{{Liu} {et~al.}(2020){Liu}, {Lambert}, {Zhu}, \& {Liu}}]{LiuN2020}
{Liu}, N., {Lambert}, S.~B., {Zhu}, Z., \& {Liu}, J.~C. 2020, \aap, 634, A28,
  \dodoi{10.1051/0004-6361/201936996}

\bibitem[{{Liu} {et~al.}(2018){Liu}, {Zhu}, \& {Liu}}]{LiuN2018}
{Liu}, N., {Zhu}, Z., \& {Liu}, J.~C. 2018, \aap, 609, A19,
  \dodoi{10.1051/0004-6361/201732006}

\bibitem[{{Ma} {et~al.}(1998){Ma}, {Arias}, {Eubanks}, {Fey}, {Gontier},
  {Jacobs}, {Sovers}, {Archinal}, \& {Charlot}}]{Ma1998}
{Ma}, C., {Arias}, E.~F., {Eubanks}, T.~M., {et~al.} 1998, \aj, 116, 516,
  \dodoi{10.1086/300408}

\bibitem[{{MacMillan} {et~al.}(2019){MacMillan}, {Fey}, {Gipson}, {Gordon},
  {Jacobs}, {Kr{\'a}sn{\'a}}, {Lambert}, {Malkin}, {Titov}, {Wang}, \&
  {Xu}}]{MacMillan2019}
{MacMillan}, D.~S., {Fey}, A., {Gipson}, J.~M., {et~al.} 2019, \aap, 630, A93,
  \dodoi{10.1051/0004-6361/201935379}

\bibitem[{{Makarov} {et~al.}(2012){Makarov}, {Dorland}, {Gaume}, {Hennessy},
  {Berghea}, {Dudik}, \& {Schmitt}}]{Makarov2012}
{Makarov}, V.~V., {Dorland}, B.~N., {Gaume}, R.~A., {et~al.} 2012, \aj, 144,
  22, \dodoi{10.1088/0004-6256/144/1/22}

\bibitem[{{Makarov} {et~al.}(2023){Makarov}, {Johnson}, \&
  {Secrest}}]{Makarov2023}
{Makarov}, V.~V., {Johnson}, M.~C., \& {Secrest}, N.~J. 2023, \aj, 166, 8,
  \dodoi{10.3847/1538-3881/acd84c}

\bibitem[{{Makarov} \& {Murphy}(2007)}]{Makarov2007}
{Makarov}, V.~V., \& {Murphy}, D.~W. 2007, \aj, 134, 367,
  \dodoi{10.1086/518242}

\bibitem[{{Malkin}(2014)}]{Malkin2014}
{Malkin}, Z. 2014, \mnras, 445, 845, \dodoi{10.1093/mnras/stu1796}

\bibitem[{{Malkin}(2018)}]{Malkin2018}
---. 2018, \apjs, 239, 20, \dodoi{10.3847/1538-4365/aae777}

\bibitem[{{Malkin}(2024{\natexlab{a}})}]{Malkin2024_ICRS}
---. 2024{\natexlab{a}}, \aj, 167, 229, \dodoi{10.3847/1538-3881/ad35bf}

\bibitem[{{Malkin}(2024{\natexlab{b}})}]{Malkin2024_VSH}
---. 2024{\natexlab{b}}, arXiv e-prints, arXiv:2410.06075,
  \dodoi{10.48550/arXiv.2410.06075}

\bibitem[{{Malkin} \& {Titov}(2008)}]{Malkin2008}
{Malkin}, Z., \& {Titov}, O. 2008, in Measuring the Future, Proceedings of the
  Fifth IVS, ed. A.~{Finkelstein} \& D.~{Behrend}, 183--187,
  \dodoi{10.48550/arXiv.0911.3216}

\bibitem[{{Marcote} {et~al.}(2018){Marcote}, {Rib{\'o}}, {Paredes}, {Mao}, \&
  {Edwards}}]{Marcote2018}
{Marcote}, B., {Rib{\'o}}, M., {Paredes}, J.~M., {Mao}, M.~Y., \& {Edwards},
  P.~G. 2018, \aap, 619, A26, \dodoi{10.1051/0004-6361/201832572}

\bibitem[{{Mayer} \& {B{\"o}hm}(2022)}]{Mayer2022}
{Mayer}, D., \& {B{\"o}hm}, J. 2022, in {Beyond 100: The Next Century in
  Geodesy}, ed. J.~T. {Freymueller} \& L.~{S{\'a}nchez} (Cham: Springer
  International Publishing), 21--28

\bibitem[{{Mignard} \& {Klioner}(2012)}]{Mignard2012}
{Mignard}, F., \& {Klioner}, S. 2012, \aap, 547, A59,
  \dodoi{10.1051/0004-6361/201219927}

\bibitem[{{Mignard} \& {Morando}(1990)}]{Mignard1990}
{Mignard}, F., \& {Morando}, B. 1990, in Journ{\'e}es 1990: Syst{\`e}mes de
  R{\'e}f{\'e}rence Spatio-Temporels, ed. N.~{Capitaine} \& S.~{D{\'e}barbat},
  151--158

\bibitem[{{Mignard} {et~al.}(2016){Mignard}, {Klioner}, {Lindegren}, {Bastian},
  {Bombrun}, {Hern{\'a}ndez}, {Hobbs}, {Lammers}, {Michalik}, {Ramos-Lerate},
  {Biermann}, {Butkevich}, {Comoretto}, {Joliet}, {Holl}, {Hutton}, {Parsons},
  {Steidelm{\"u}ller}, {Andrei}, {Bourda}, \& {Charlot}}]{Mignard2016}
{Mignard}, F., {Klioner}, S., {Lindegren}, L., {et~al.} 2016, \aap, 595, A5,
  \dodoi{10.1051/0004-6361/201629534}

\bibitem[{{Momjian} {et~al.}(2021){Momjian}, {Ba{\~n}ados}, {Carilli},
  {Walter}, \& {Mazzucchelli}}]{Momjian2021}
{Momjian}, E., {Ba{\~n}ados}, E., {Carilli}, C.~L., {Walter}, F., \&
  {Mazzucchelli}, C. 2021, \aj, 161, 207, \dodoi{10.3847/1538-3881/abe6ae}

\bibitem[{{Momjian} {et~al.}(2004){Momjian}, {Petric}, \&
  {Carilli}}]{Momjian2004}
{Momjian}, E., {Petric}, A.~O., \& {Carilli}, C.~L. 2004, \aj, 127, 587,
  \dodoi{10.1086/381300}

\bibitem[{{Nothnagel} {et~al.}(2017){Nothnagel}, {Artz}, {Behrend}, \&
  {Malkin}}]{Nothnagel2017}
{Nothnagel}, A., {Artz}, T., {Behrend}, D., \& {Malkin}, Z. 2017, Journal of
  Geodesy, 91, 711, \dodoi{10.1007/s00190-016-0950-5}

\bibitem[{{Petrov}(2011)}]{Petrov2011_OBRS1}
{Petrov}, L. 2011, \aj, 142, 105, \dodoi{10.1088/0004-6256/142/4/105}

\bibitem[{{Petrov}(2012)}]{Petrov2012_EGaPS}
---. 2012, \mnras, 419, 1097, \dodoi{10.1111/j.1365-2966.2011.19765.x}

\bibitem[{{Petrov}(2013)}]{Petrov2013_OBRS2}
---. 2013, \aj, 146, 5, \dodoi{10.1088/0004-6256/146/1/5}

\bibitem[{{Petrov}(2021)}]{Petrov2021_WFCS}
---. 2021, \aj, 161, 14, \dodoi{10.3847/1538-3881/abc4e1}

\bibitem[{{Petrov} {et~al.}(2019){Petrov}, {de Witt}, {Sadler}, {Phillips}, \&
  {Horiuchi}}]{Petrov2019_LCS2}
{Petrov}, L., {de Witt}, A., {Sadler}, E.~M., {Phillips}, C., \& {Horiuchi}, S.
  2019, \mnras, 485, 88, \dodoi{10.1093/mnras/stz242}

\bibitem[{{Petrov} {et~al.}(2011{\natexlab{a}}){Petrov}, {Kovalev}, {Fomalont},
  \& {Gordon}}]{Petrov2011_VGaPS}
{Petrov}, L., {Kovalev}, Y.~Y., {Fomalont}, E.~B., \& {Gordon}, D.
  2011{\natexlab{a}}, \aj, 142, 35, \dodoi{10.1088/0004-6256/142/2/35}

\bibitem[{{Petrov} {et~al.}(2011{\natexlab{b}}){Petrov}, {Phillips},
  {Bertarini}, {Murphy}, \& {Sadler}}]{Petrov2011_LCS1}
{Petrov}, L., {Phillips}, C., {Bertarini}, A., {Murphy}, T., \& {Sadler}, E.~M.
  2011{\natexlab{b}}, \mnras, 414, 2528,
  \dodoi{10.1111/j.1365-2966.2011.18570.x}

\bibitem[{{Petrov} \& {Taylor}(2011)}]{Petrov2011_VIPS}
{Petrov}, L., \& {Taylor}, G.~B. 2011, \aj, 142, 89,
  \dodoi{10.1088/0004-6256/142/3/89}

\bibitem[{{Petrov} {et~al.}(2012){Petrov}, {Lee}, {Kim}, {Jung}, {Oh}, {Sohn},
  {Byun}, {Chung}, {Je}, {Wi}, {Song}, {Kang}, {Han}, {Lee}, {Kim}, {Chung}, \&
  {Kim}}]{Petrov2012_QCAL1}
{Petrov}, L., {Lee}, S.-S., {Kim}, J., {et~al.} 2012, \aj, 144, 150,
  \dodoi{10.1088/0004-6256/144/5/150}

\bibitem[{{Petrov} \& {Kovalev}(2025)}]{Petrov2025_RFC}
{Petrov}, L.~Y., \& {Kovalev}, Y.~Y. 2025, \apjs, 276, 38,
  \dodoi{10.3847/1538-4365/ad8c36}

\bibitem[{{Popkov} {et~al.}(2021){Popkov}, {Kovalev}, {Petrov}, \&
  {Kovalev}}]{Popkov2021}
{Popkov}, A.~V., {Kovalev}, Y.~Y., {Petrov}, L.~Y., \& {Kovalev}, Y.~A. 2021,
  \aj, 161, 88, \dodoi{10.3847/1538-3881/abd18c}

\bibitem[{{Schinzel} {et~al.}(2015){Schinzel}, {Petrov}, {Taylor}, {Mahony},
  {Edwards}, \& {Kovalev}}]{Schinzel2015_2FGL}
{Schinzel}, F.~K., {Petrov}, L., {Taylor}, G.~B., {et~al.} 2015, \apjs, 217, 4,
  \dodoi{10.1088/0067-0049/217/1/4}

\bibitem[{{Shu} {et~al.}(2017){Shu}, {Petrov}, {Jiang}, {Xia}, {Jiang}, {Cui},
  {Takefuji}, {McCallum}, {Lovell}, {Yi}, {Hao}, {Yang}, {Zhang}, {Chen}, \&
  {Li}}]{Shu2017_VEPS1}
{Shu}, F., {Petrov}, L., {Jiang}, W., {et~al.} 2017, \apjs, 230, 13,
  \dodoi{10.3847/1538-4365/aa71a3}

\bibitem[{{Titov} \& {Lambert}(2013)}]{Titov2013}
{Titov}, O., \& {Lambert}, S. 2013, \aap, 559, A95,
  \dodoi{10.1051/0004-6361/201321806}

\bibitem[{{Titov} {et~al.}(2011){Titov}, {Lambert}, \& {Gontier}}]{Titov2011}
{Titov}, O., {Lambert}, S.~B., \& {Gontier}, A.~M. 2011, \aap, 529, A91,
  \dodoi{10.1051/0004-6361/201015718}

\bibitem[{{Vityazev} \& {Shuksto}(2004)}]{Vityazev2004}
{Vityazev}, V., \& {Shuksto}, A. 2004, in Astronomical Society of the Pacific
  Conference Series, Vol. 316, Order and Chaos in Stellar and Planetary
  Systems, ed. G.~G. {Byrd}, K.~V. {Kholshevnikov}, A.~A. {Myllri}, I.~I.
  {Nikiforov}, \& V.~V. {Orlov}, 230--233

\bibitem[{{Vityazev}(2017)}]{Vityazev2017B}
{Vityazev}, V.~V. 2017, {Analysis of astrometric catalogs using spherical
  functions} ({St. Petersburg State University, Russia (in Russian)})

\bibitem[{{Winn} {et~al.}(2002){Winn}, {Lovell}, {Chen}, {Fletcher}, {Hewitt},
  {Patnaik}, \& {Schechter}}]{Winn2002}
{Winn}, J.~N., {Lovell}, J. E.~J., {Chen}, H.-W., {et~al.} 2002, \apj, 564,
  143, \dodoi{10.1086/324144}

\bibitem[{{Yang} {et~al.}(2012){Yang}, {Wu}, {Paragi}, \& {An}}]{Yang2012}
{Yang}, J., {Wu}, F., {Paragi}, Z., \& {An}, T. 2012, \mnras, 419, L74,
  \dodoi{10.1111/j.1745-3933.2011.01182.x}

\bibitem[{{Yao} {et~al.}(2024){Yao}, {Liu}, {Liu}, {Zhu}, \& {Wang}}]{Yao2024}
{Yao}, J., {Liu}, J.-C., {Liu}, N., {Zhu}, Z., \& {Wang}, Z.-W. 2024, Research
  in Astronomy and Astrophysics, 24, 085011, \dodoi{10.1088/1674-4527/ad621e}

\end{thebibliography}
\bibliographystyle{aasjournal}

%%%%%%%%%%%%%%%%%%%%%%%%%%%%%%%%%%%%%%%%%%%%%%%%%%%%%%%%%%%%%%%%%

\appendix
\section*{Explicit expressions for VSH decomposition used in this work}

Vector spherical harmonics (VSH) are widely used in astronomical
research to analyze vector fields on the celestial sphere.
In practice, the problem of decompositing the analyzed vector field
$\ve{V}(\alpha,\delta)$ into VSH is solved by determination of
coefficients $a_i$ by the LS method from the following system
of equations:
\begin{equation}
\ve{V}(\alpha_j,\delta_j) = \sum_{i=1}^{N_{coeff}} a_i \ve{F}_i(\alpha_j,\delta_j) \,,
\label{eq:vshlsp2}
\end{equation}
where j~--the object number (star, extragalactic radio source)
or cell number if a pixelizaton is used, $\ve{F}$~-- VSH,
$N_{coeff}$~-- the number of coefficients, which depends on
the degree of the VSH decomposition $l$ as $N_{coeff}$=$2l(l+2)$.

Different authors implement slightly different versions of the VSH
decomposition and use different notations of harmonics
\citep{Mignard1990,Vityazev2004,Makarov2007,Mignard2012,Vityazev2017B},
which are compared in \cite{Malkin2024_VSH}.
In this paper, the 48-term decompositions ($l$=4) as given by
\cite[Appendix A]{Mignard2012} is used.
Corresponding VSH coefficients to be adjusted in solution are explicitly
listed in Table~\ref{tab:vsh4_my}.
In the last column of the Table, a correspondence is shown between
the notation of \cite{Mignard2012} and the notation accepted
in a 16-term expansion up to $l$=2 most widely used in ICRF-related papers
\citep{Titov2013,LiuN2018,Karbon2019,Charlot2020,LiuN2020,Mayer2022,Krasna2023,Yao2024},
which is given by:
\begin{equation}
\begin{array}{rcl}
\Delta\alpha^{\ast} & = &\phantom{-}R_1 \cos\alpha \sin\delta + R_2 \sin\alpha \sin\delta - R_3 \cos\delta \\[0.5ex]
                    &   &- \, G_1 \sin\alpha + G_2 \cos\alpha \\[0.5ex]
                    &   &+ \, M_{2,0} \sin 2\delta \\[0.7ex]
                    &   &- \, (M_{2,1}^{\rm Re} \cos\alpha - M_{2,1}^{\rm Im} \sin\alpha) \cos 2\delta \\[1ex]
                    &   &+ \, (E_{2,1}^{\rm Re} \sin\alpha + E_{2,1}^{\rm Im} \cos\alpha) \sin\delta \\[1ex]
                    &   &- \, (M_{2,2}^{\rm Re} \cos 2\alpha - M_{2,2}^{\rm Im} \sin 2\alpha) \sin 2\delta \\[1ex]
                    &   &- \, 2 \, (E_{2,2}^{\rm Re} \sin 2\alpha + E_{2,2}^{\rm Im} \cos 2\alpha) \cos\delta \,, \\[2ex]
\Delta\delta        & = &-R_1 \sin\alpha + R_2 \cos\alpha \\[0.5ex]
                    &   &- \, G_1 \cos\alpha \sin\delta - G_2 \sin\alpha \sin\delta + G_3 \cos\delta \\[0.5ex]
                    &   &+ \, E_{2,0} \sin 2\delta  \\[0.7ex]
                    &   &- \, (M_{2,1}^{\rm Re} \sin\alpha + M_{2,1}^{\rm Im} \cos\alpha) \sin\delta \\[1ex]
                    &   &- \, (E_{2,1}^{\rm Re} \cos\alpha - E_{2,1}^{\rm Im} \sin\alpha) \cos 2\delta \\[1ex]
                    &   &+ \, 2 \, (M_{2,2}^{\rm Re} \sin 2\alpha + M_{2,2}^{\rm Im} \cos 2\alpha) \cos\delta \\[1ex]
                    &   &- \, (E_{2,2}^{\rm Re} \cos 2\alpha - E_{2,2}^{\rm Im} \sin 2\alpha) \sin 2\delta \,, \\[1ex]
\end{array}
\label{eq:vsh2}
\end{equation}
where $\Delta\alpha^* = \Delta\alpha\cos\delta$.
The first degree harmonics consists of a rotation vector $\mathbf{R} = (R_1, R_2, R_3)^{\mathrm{T}}$
and a glide vector $\mathbf{G} = (G_1, G_2, G_3)^{\mathrm{T}}$.
The rotation vector $\mathbf{R}$ characterizes the mutual orientation between frames.
The glide vector $\mathbf{G}$ reflects the dipolar deformation in the celestial frame.
The second degree terms are also called quadrupole terms.

As can be seen from these examples and Table~\ref{tab:vsh4_my}, in this work,
the model coefficients are adjusted directly for trigonometric functions without
taking into account the normalizing factor.
This approach can be justified by a direct correspondence between the found
coefficients and the parameters of a physical phenomenon, such as the mutual
rotation between catalog systems or the kinematic parameters of the Galaxy.
The intra-source correlations between R.A. and decl. were used in computations.

\clearpage
\begin{table*}%[hb]
\centering
\caption{Toroidal (\ve{T}) and spheroidal (\ve{S}) VSH functions.
  L2 column shows designations of the coefficients in Eq.~(\ref{eq:vsh2}).}
\label{tab:vsh4_my}
\begin{tabular}{ccccc}
\hline
No. & \ve{F} & $\Delta\alpha^{\ast}$ & $\Delta\delta$ & L2 \\
\hline \\[-6pt]
 1 & $\ve{T}_{10}$          & $\cos\delta$                                & ---                       & $-R_3$             \\[1ex]
 2 & $\ve{S}_{10}$          & ---                                         & $\cos\delta$              & $~G_3$             \\[1ex]
 3 & $\ve{T}_{11}^{\rm Re}$ & $\cos\alpha\sin\delta$                      & $-\sin\alpha$             & $~R_1$             \\[1ex]
 4 & $\ve{T}_{11}^{\rm Im}$ & $\sin\alpha\sin\delta$                      & $ \cos\alpha$             & $~R_2$             \\[1ex]
 5 & $\ve{S}_{11}^{\rm Re}$ & $ \sin\alpha$                               & $\cos\alpha\sin\delta$    & $-G_1$             \\[1ex]
 6 & $\ve{S}_{11}^{\rm Im}$ & $-\cos\alpha$                               & $\sin\alpha\sin\delta$    & $-G_2$             \\[1ex]
 7 & $\ve{T}_{20}$          & $\sin2\delta$                               & ---                       & $M_{20}$           \\[1ex]
 8 & $\ve{S}_{20}$          & ---                                         & $\sin2\delta$             & $E_{20}$           \\[1ex]
 9 & $\ve{T}_{21}^{\rm Re}$ & $-\cos\alpha\cos2\delta$                    & $-\sin\alpha\sin\delta$   & $~M_{21}^{\rm Re}$ \\[1ex]
10 & $\ve{T}_{21}^{\rm Im}$ & $-\sin\alpha\cos2\delta$                    & $ \cos\alpha\sin\delta$   & $-M_{21}^{\rm Im}$ \\[1ex]
11 & $\ve{S}_{21}^{\rm Re}$ & $ \sin\alpha\sin\delta$                     & $-\cos\alpha\cos2\delta$  & $~E_{21}^{\rm Re}$ \\[1ex]
12 & $\ve{S}_{21}^{\rm Im}$ & $-\cos\alpha\sin\delta$                     & $-\sin\alpha\cos2\delta$  & $-E_{21}^{\rm Im}$ \\[1ex]
13 & $\ve{T}_{22}^{\rm Re}$ & $-\cos2\alpha\sin2\delta$                   & $ 2\sin2\alpha\cos\delta$ & $~M_{22}^{\rm Re}$ \\[1ex]
14 & $\ve{T}_{22}^{\rm Im}$ & $-\sin2\alpha\sin2\delta$                   & $-2\cos2\alpha\cos\delta$ & $-M_{22}^{\rm Im}$ \\[1ex]
15 & $\ve{S}_{22}^{\rm Re}$ & $-2\sin2\alpha\cos\delta$                   & $-\cos2\alpha\sin2\delta$ & $~E_{22}^{\rm Re}$ \\[1ex]
16 & $\ve{S}_{22}^{\rm Im}$ & $ 2\cos2\alpha\cos\delta$                   & $-\sin2\alpha\sin2\delta$ & $-E_{22}^{\rm Im}$ \\[1ex]
17 & $\ve{T}_{30}$          & $\cos\delta(5\sin^2\!\delta-1)$             & ---                                         &   \\[1ex]
18 & $\ve{S}_{30}$          & ---                                         & $\cos\delta(5\sin^2\!\delta-1)$             &   \\[1ex]
19 & $\ve{T}_{31}^{\rm Re}$ & $\cos\alpha\sin\delta(15\sin^2\!\delta-11)$ & $-\sin\alpha(5\sin^2\!\delta-1)$            &   \\[1ex]
20 & $\ve{T}_{31}^{\rm Im}$ & $\sin\alpha\sin\delta(15\sin^2\!\delta-11)$ & $ \cos\alpha(5\sin^2\!\delta-1)$            &   \\[1ex]
21 & $\ve{S}_{31}^{\rm Re}$ & $ \sin\alpha(5\sin^2\!\delta-1)$            & $\cos\alpha\sin\delta(15\sin^2\!\delta-11)$ &   \\[1ex]
22 & $\ve{S}_{31}^{\rm Im}$ & $-\cos\alpha(5\sin^2\!\delta-1)$            & $\sin\alpha\sin\delta(15\sin^2\!\delta-11)$ &   \\[1ex]
23 & $\ve{T}_{32}^{\rm Re}$ & $-\cos2\alpha\cos\delta(3\sin^2\!\delta-1)$ & $ \sin2\alpha\sin2\delta$                   &   \\[1ex]
24 & $\ve{T}_{32}^{\rm Im}$ & $-\sin2\alpha\cos\delta(3\sin^2\!\delta-1)$ & $-\cos2\alpha\sin2\delta$                   &   \\[1ex]
25 & $\ve{S}_{32}^{\rm Re}$ & $-\sin2\alpha\sin2\delta$                   & $-\cos2\alpha\cos\delta(3\sin^2\!\delta-1)$ &   \\[1ex]
26 & $\ve{S}_{32}^{\rm Im}$ & $ \cos2\alpha\sin2\delta$                   & $-\sin2\alpha\cos\delta(3\sin^2\!\delta-1)$ &   \\[1ex]
27 & $\ve{T}_{33}^{\rm Re}$ & $\cos3\alpha\cos^2\!\delta\sin\delta$       & $-\sin3\alpha\cos^2\!\delta$                &   \\[1ex]
28 & $\ve{T}_{33}^{\rm Im}$ & $\sin3\alpha\cos^2\!\delta\sin\delta$       & $ \cos3\alpha\cos^2\!\delta$                &   \\[1ex]
29 & $\ve{S}_{33}^{\rm Re}$ & $ \sin3\alpha\cos^2\!\delta$                & $\cos3\alpha\cos^2\!\delta\sin\delta$       &   \\[1ex]
30 & $\ve{S}_{33}^{\rm Im}$ & $-\cos3\alpha\cos^2\!\delta$                & $\sin3\alpha\cos^2\!\delta\sin\delta$       &   \\[1ex]
\hline
\end{tabular}
\end{table*}

\addtocounter{table}{-1}
\begin{table*}
\centering
\caption{(continue) Toroidal (\ve{T}) and spheroidal (\ve{S}) VSH functions.}
\begin{tabular}{cccc}
\hline
No. & \ve{F} & $\Delta\alpha^{\ast}$ & $\Delta\delta$ \\
\hline \\[-6pt]
31 & $\ve{T}_{40}$          & $\sin2\delta(7\sin^2\!\delta-3)$                  & ---                                               \\[1ex]
32 & $\ve{S}_{40}$          & ---                                               & $\sin2\delta(7\sin^2\!\delta-3)$                  \\[1ex]
33 & $\ve{T}_{41}^{\rm Re}$ & $\cos\alpha(28\sin^4\!\delta-27\sin^2\!\delta+3)$ & $-\sin\alpha\sin\delta(7\sin^2\!\delta-3)$        \\[1ex]
34 & $\ve{T}_{41}^{\rm Im}$ & $\sin\alpha(28\sin^4\!\delta-27\sin^2\!\delta+3)$ & $ \cos\alpha\sin\delta(7\sin^2\!\delta-3)$        \\[1ex]
35 & $\ve{S}_{41}^{\rm Re}$ & $ \sin\alpha\sin\delta(7\sin^2\!\delta-3)$        & $\cos\alpha(28\sin^4\!\delta-27\sin^2\!\delta+3)$ \\[1ex]
36 & $\ve{S}_{41}^{\rm Im}$ & $-\cos\alpha\sin\delta(7\sin^2\!\delta-3)$        & $\sin\alpha(28\sin^4\!\delta-27\sin^2\!\delta+3)$ \\[1ex]
37 & $\ve{T}_{42}^{\rm Re}$ & $-\cos2\alpha\sin2\delta(7\sin^2\!\delta-4)$      & $ \sin2\alpha\cos\delta(7\sin^2\!\delta-1)$       \\[1ex]
38 & $\ve{T}_{42}^{\rm Im}$ & $-\sin2\alpha\sin2\delta(7\sin^2\!\delta-4)$      & $-\cos2\alpha\cos\delta(7\sin^2\!\delta-1)$       \\[1ex]
39 & $\ve{S}_{42}^{\rm Re}$ & $-\sin2\alpha\cos\delta(7\sin^2\!\delta-1)$       & $-\cos2\alpha\sin2\delta(7\sin^2\!\delta-4)$      \\[1ex]
40 & $\ve{S}_{42}^{\rm Im}$ & $ \cos2\alpha\cos\delta(7\sin^2\!\delta-1)$       & $-\sin2\alpha\sin2\delta(7\sin^2\!\delta-4)$      \\[1ex]
41 & $\ve{T}_{43}^{\rm Re}$ & $\cos3\alpha\cos^2\!\delta(4\sin^2\!\delta-1)$    & $-3\sin3\alpha\cos^2\!\delta\sin\delta$           \\[1ex]
42 & $\ve{T}_{43}^{\rm Im}$ & $\sin3\alpha\cos^2\!\delta(4\sin^2\!\delta-1)$    & $~3\cos3\alpha\cos^2\!\delta\sin\delta$           \\[1ex]
43 & $\ve{S}_{43}^{\rm Re}$ & $ 3\sin3\alpha\cos^2\!\delta\sin\delta$           & $\cos3\alpha\cos^2\!\delta(4\sin^2\!\delta-1)$    \\[1ex]
44 & $\ve{S}_{43}^{\rm Im}$ & $-3\cos3\alpha\cos^2\!\delta\sin\delta$           & $\sin3\alpha\cos^2\!\delta(4\sin^2\!\delta-1)$    \\[1ex]
45 & $\ve{T}_{44}^{\rm Re}$ & $-\cos4\alpha\cos^3\!\delta\sin\delta$            & $ \sin4\alpha\cos^3\!\delta$                      \\[1ex]
46 & $\ve{T}_{44}^{\rm Im}$ & $-\sin4\alpha\cos^3\!\delta\sin\delta$            & $-\cos4\alpha\cos^3\!\delta$                      \\[1ex]
47 & $\ve{S}_{44}^{\rm Re}$ & $-\sin4\alpha\cos^3\!\delta$                      & $-\cos4\alpha\cos^3\!\delta\sin\delta$            \\[1ex]
48 & $\ve{S}_{44}^{\rm Im}$ & $ \cos4\alpha\cos^3\!\delta$                      & $-\sin4\alpha\cos^3\!\delta\sin\delta$            \\[1ex]
\hline
\end{tabular}
\end{table*}

\end{document}